\documentclass[copyright,creativecommons]{eptcs}
\providecommand{\event}{FORECAST 2016}

\usepackage{graphicx}
\usepackage{caption}
\usepackage{subcaption}

\title{From Collective Adaptive Systems to \\ Human Centric Computation and Back: \\ Spatial Model Checking for Medical Imaging 
	}
\author{Gina Belmonte\thanks{Research partially supported by the Azienda Ospedaliera Universitaria Senese}
\institute{Azienda Ospedaliera Universitaria Senese\\ Siena, Italy}
\email{g.belmonte@ao-siena.toscana.it}
\and
Vincenzo Ciancia\thanks{Research partially supported by the EU project QUANTICOL (nr. 600708)} \qquad Diego Latella\footnotemark[2] \qquad Mieke Massink\footnotemark[2]
\institute{Istituto di Scienza e Tecnologie dell'Informazione ``A. Faedo''\\
	Consiglio Nazionale delle Ricerche \\ Pisa, Italy}
\email{\quad \{vincenzo.ciancia, diego.latella, mieke.massink\}@isti.cnr.it}
}
\def\titlerunning{Spatial Model Checking for Medical Imaging}
\def\authorrunning{G. Belmonte, V. Ciancia , D. Latella, M. Massink}

	\usepackage{amsmath}
\usepackage{amssymb}
\usepackage{xspace}

\newcommand{\eval}{v}
\newcommand{\props}{P}
\newcommand{\model}{\mathcal{M}}

\newcommand{\closure}{\mathcal{C}}
\newcommand{\dop}{d}

\newcommand{\dist}[1]{\mathcal{D}^{#1}}

\newcommand{\reals}{\mathbb{R}}

\newcommand{\arel}{\mathcal{R}}

\newcommand{\topochecker}{{\ttfamily{topochecker}}\xspace}
\newcommand{\SLCS}{SLCS\xspace}

\usepackage{amsmath}
\usepackage{amssymb}

\newcommand{\lnear}{\mathcal{N}}
\newcommand{\lsurr}{\mathcal{S}}
\newcommand{\pow}{\mathcal{P}}

\usepackage{xcolor}
\hypersetup{%
	colorlinks=false,% hyperlinks will be black
	linkbordercolor=orange,% hyperlink borders will be red
	pdfborderstyle={/S/U/W .4}% border style will be underline of width 1pt
}
\begin{document}
\maketitle

\begin{abstract}
Recent research on formal verification for Collective Adaptive Systems (CAS) pushed advancements in spatial and spatio-temporal model checking, and as a side result  provided novel image analysis methodologies, rooted in logical methods for topological spaces. Medical Imaging (MI) is a field where such technologies show potential for ground-breaking innovation. In this position paper, we present a preliminary investigation centred on applications of spatial model checking to MI. The focus is shifted from pure logics to a mixture of logical, statistical and algorithmic approaches, driven by the logical nature intrinsic to the specification of the properties of interest in the field. As a result, novel operators are introduced, that could as well be brought back to the setting of CAS.
\end{abstract}
%!TEX root=main.tex

\section{Introduction}
Formal verification of properties of Collective Adaptive Systems (CAS)
is a challenging subject. The huge number of considered entities 
introduces a gap with classical finite-state methods as the
number of states grows exponentially. Approximation methods, such as mean-field or fluid-flow approximation, have been
proposed to mitigate this aspect (see \cite{BH12,BHLM13,LLM15}). Another relevant issue is that of
spatial distribution of the considered entities. Entities composing a CAS are typically located, and moving, in a physical or logical space. Collective behaviour is driven by interaction, which is frequently based on proximity. This makes spatial aspects more prominent in the case of CAS than in classical concurrent systems, and leads to the
introduction of spatial properties in formal verification.

Model checking \cite{BK08} is a formal verification technique that is based on
static analysis of system properties that are described by
\emph{modal} logics. Modal operators have been
traditionally used to denote possibility or necessity,
probability, constraints on continuous time, access to security
contexts, separation of parallel components of a system and so on. But
since the very beginning, modal logics have also been interpreted on
\emph{spatial} structures, such as topological spaces (see
\cite{Ch5HBSL} for a thorough introduction). In this context, formulas
are interpreted as sets of points of a topological space, and in
particular $\diamond \phi$ is usually interpreted as the points that
lay in the \emph{closure} of the interpretation of $\phi$.  A standard reference is the \emph{Handbook of Spatial Logics} 
\cite{HBSL}. Therein, several logics are described, with applications far
beyond topological spaces; such logics treat not only aspects of
morphology, geometry, distance, but also advanced topics such as
dynamic systems, and discrete structures, that are particularly
difficult to deal with from a topological perspective.

Model checking of spatial (and spatio-temporal) logics is a more
recent development (see e.g., \cite{De+07}, \cite{Gr+09}, \cite{SPATEL}). In \cite{CLLM14}, a model checking algorithm for spatial logics in the topological sense has been proposed. Therein, models are based on extensions of topological spaces called \emph{closure spaces}, designed to accommodate also discrete structures such as finite graphs in the topological framework. The so-called \emph{surrounded} operator is introduced to describe points that are located in a region of points
satisfying a certain formula and surrounded by
points satisfying another one (in other words, the surrounded operator is a spatial form of the
\emph{until} operator of temporal logics). The resulting logic is called \SLCS (Spatial Logic of Closure Spaces). In \cite{CGLLM15}, a
\emph{spatio-temporal} model checking algorithm using the same principles has
been proposed, for a logic combining the spatial operators of SLCS with the well-known temporal operators of the \emph{Computation Tree Logic} CTL. Applications to CAS have been developed in the fields
of \emph{smart transportation} \cite{CGLLM14}, and \emph{bike-sharing
  systems} \cite{CLMP15}. A free and open source implementation of the spatio-temporal model checking algorithm of \cite{CGLLM15} is provided by the tool \topochecker\footnote{See
  \url{http://topochecker.isti.cnr.it/} and
  \url{https://github.com/vincenzoml/topochecker}.}.

In this position paper, we explore various ideas for the application of spatial and spatio-temporal
model checking that depart from the setting of CAS and are tailored to
Human Centric Computation, in particular to the field of \emph{Medical
  Imaging} (MI). In this domain, space consists of points
called \emph{voxels}, that are arranged in a multi-dimensional, possibly anisotropic
grid. Several spatial analyses are performed in
MI. Such analyses are typically described by structured combinations of attributes related to proximity, shape, aspect and distance of features of interest, for which spatial logics provide a well-suited descriptive language. 
Our work is motivated by some considerations about how medical image analysis is carried out. The overall general description of a feature (e.g. the shape and spatial arrangement of parts of an image that exhibit diseases) is often carried out informally, but in a logically structured way (e.g.: ``the tumour is lighter than the surrounding brain area, and touches the oedema, whose intensity is a bit darker than the tumour''). Such description is then turned into a series of different analysis passes, sometimes performed by specific tools, but frequently done by writing ad-hoc programs. The results of such different passes are often integrated by hand or using hand-crafted scripts. This complex and elaborate process hinders the implementation, and sharing across the medical community, of novel analysis methods that emerge from current research. A leap forward is needed in the unambiguous and precise specification of such procedures, which could be provided by logical methods borrowed from Computer Science and in particular the area of \emph{formal methods}. Our research program aims at paving the way and establishing foundational results for such a development to happen.

%!TEX root=main.tex

\section{Spatial logics for medical imaging}

% \begin{itemize}
% 	\item Why is it important to give a logical structure to medical imaging (see the proposal)
% 	\item What are the most relevant applications (what is contouring? what is CAD? pointers to related work (e.g. cite a literature review))
% 	\item Discuss the temporal aspects (e.g. the $b$ parameter in MRI?, or multiple acquisitions?)
% \end{itemize}

The contribution of Computer Science to the field of medical image
analysis is increasingly significant, and will play a key role in
future healthcare.  Computational methods are currently in use for
several different purposes, such as: 
%\begin{enumerate}
%\item 
\emph{Computer-Aided Diagnosis} (CAD), aiming at the
  classification of areas in images, based on the presence of signs of
  specific diseases \cite{Doi2007};
%\item 
\emph{Image Segmentation}, tailored to identify areas that
  exhibit specific features or functions (e.g. organs or
  sub-structures) \cite{Gordillo2013};
%\item 
\emph{Automatic contouring} of Organs at Risk (OAR) or target
  volumes (TV) for radiotherapy applications \cite{brock2014image};
%\item 
\emph{Indicators finding}, that is, the identification of indicators, computed from the acquired images, 
	that permit early diagnosis, or understanding of microscopic characteristics of specific
  diseases, or help in the identification of prognostic factors to predict a treatment
  output \cite{Chetelat2003, Toosy2003}. Examples of indicators are
  the Mean Diffusivity and the Fractional Anisotropy obtained from 
  Magnetic Resonance (MR) Diffusion-Weighted Images, or 
  Magnetization Transfer Ratio maps obtained from a Magnetization
  Transfer acquisition \cite{DeSantis2014, Li2015}.
%\end{enumerate}

Such kinds of analyses are strictly related to spatial and temporal features
of acquired images. For example, the diagnosis of specific diseases
requires the observation of variations in time of the response to a
particular acquisition technique. Another example is provided by
\emph{longitudinal studies}, that consist of repeated observations of
the same variables over long periods of time, to help understanding
the areas involved in some diseases.  Such investigations can take
advantage of spatial and temporal capabilities in model checking. In
this work, we provide a preliminary investigation of spatial logical operators
that may be used to identify areas in the images based on local or
non-local features and prior knowledge (e.g., areas \emph{surrounded
  by} or \emph{near to} or \emph{similar to} areas with particular
properties).

We start from the spatial logic \SLCS presented in \cite{CLLM14}. \SLCS features the spatial
operators \emph{near} and \emph{surrounded}. 
The syntax of \SLCS is defined by the following grammar, where $p$ ranges over $P$, namely the set of \emph{atomic propositions}:
 	$
 	\Phi ::= 
 	p \mid \top \mid \lnot \Phi \mid \Phi \land \Phi \mid \lnear \Phi \mid \Phi \lsurr \Phi 
 	$.
 
In the syntax, $\top$ is the constant \emph{true}; $\lnot$ and $\land$ are standard logical negation and conjunction.  Formula $\lnear \phi$ denotes all points that are ``near'' to the interpretation of $\phi$, whereas formula $\phi_1 \lsurr \phi_2$ denotes all points that lie in a subset of the interpretation of $\phi_1$, whose ``boundary'' satisfies $\phi_2$. The notions of ``near'' and ``boundary'' are made formal in the interpretation of the logic, resorting to a \emph{closure model} $((X,\closure),\eval)$, where $X$ is a set, $\closure: \pow(X) \to \pow(X)$ is a \emph{closure operator} (see \cite{CLLM14} for details), and $\eval : P \to \pow(X)$ is the valuation of atomic propositions. 
 
 In the remainder of this paper, we discuss some additional logical operators that deal with distance
 (\autoref{sec:dist}) and texture analysis (\autoref{sec:ta}), and we reconsider the models of \SLCS in the light of such additions. These operators are implemented in the experimental branch of \topochecker. In models associated to medical images, $X$ is the set of points of an image, and $\closure$ associates to each subset $A$ of $X$ the union of the neighbours of each point of $A$, by a user-defined notion of neighbourhood. To cope with the quantitative information present in medical images, atomic propositions have quantitative valuations over the real numbers, instead of just boolean valuations. However, to retain the boolean interpretation of \SLCS formulas,  the syntax of atomic propositions in the logic is that of constraints with variables ranging over $\reals$. Atomic predicate $p$ is a shorthand for $p=1$. For example, formula $p > k \land q < h$ denotes all points $x \in X$ such that the value of $p$ is greater than $k$ and the value of $q$ is less than $h$. Formally, this does not require changes to \SLCS and its semantics: given a set $\hat P$ of proposition letters, with quantitative valuation $\hat \eval : \hat P \to \reals^X$, the set of atomic propositions $P$ is given by the set of all possible constraints on $\hat P$, and the valuation function $\eval : P \to 2^X$ is just evaluation of constraints, which makes use of $\hat \eval$.

%!TEX root=main.tex

\section{Distance operators}
\label{sec:dist}

Distance operators can be added to spatial logics in various ways (see
\cite{Ch9HBSL} for an introduction). Distances are very often expressed using the
real numbers $\reals$. Typically, one considers operators of the form
$\dist{e(z)} \phi$ where $e(z)$ is a constraint parametrised by a free
variable $z$, and $\phi$ is a formula denoting a spatial property. The intended semantics
is that point $x$ is a model of $\dist{e(z)}\phi$ if and only if there is
a point $y$ satisfying $\phi$ such that the \emph{distance} $d$ from
$x$ to $y$ satisfies the constraint $e(d)$. Logics of metric spaces have been introduced in \cite{KWSSZ03}; therein, the constraint $e(z)$ can only be in the form
$z \leq k$, where $k \in \reals$ (\emph{distance at most
	$k$}), or $k_1 \leq z \leq k_2$ (distance included between $k_1$ and $k_2$)\footnote{See also  \cite{NB14,NBCLM15} for examples of application of such connectives in spatio-temporal signal analysis.}. The latter is called ``doughnut operator'' in
\cite{Ch9HBSL}. Notably, the doughnut operator cannot be
expressed just using $z \leq k$ in combination with boolean operators. In the following, we
will discuss a general model checking procedure which is able to
verify the satisfaction of such formulas in an efficient way. %We will use the abbreviation
%$\dist{\leq k}$ to denote $\dist{z \leq k}$.

\paragraph{Models.} First, we need to discuss appropriate
\emph{models} for our logics. The \emph{quasi-discrete closure models}
of \cite{CLLM14} provide a starting point. A quasi-discrete closure
model is completely described by a pair $(X,\arel)$ where $\arel $ is
a binary relation on a set of points $X$ -- that is, a (possibly directed, unlabelled) graph -- and
by a valuation $\eval$ associating to each atomic proposition, in a
finite set $\props$, a set of points of $X$. This data uniquely
defines a model $\model = ((X, \closure), \eval)$ where
$\closure : 2^X \to 2^X$ is the \emph{closure} operator, that coincides with the \emph{dilation} operation of the graph
$(X,\arel)$. Such models do not include information on distances. One
possibility is to enrich the structure of $\model$ by adding a
\emph{distance} operator $\dop : X \times X \to \reals_{\geq 0}$. In medical imaging, the structure of a \emph{metric space} is a very natural setting, as typical distances are based on either Euclidean spaces or symmetric graphs. See \autoref{sec:app} for more details on the possible choices for a specific metric in medical imaging applications.
An open question is what are the additional axioms (one might say ``compatibility
conditions'') linking closure and distance. The link is well-known for
the case of topological spaces. More precisely, topological spaces are obtained from metric spaces
by defining open sets as \emph{those sets $S$ such that all points of $S$
have a neighbourhood\footnote{In this definition, \emph{neighbourhood} has to be interpreted in the context of metric spaces; namely, a set $S$ is a neighbourhood of a point $x$ whenever there is $r\in \reals$ such that the set $\{y \mid \dop(x,y) < r \}$ is contained in $S$.} in $S$}; a generalization to closure spaces is an interesting topic for future research.
 
\paragraph{Distances.} In the case of quasi-discrete closure spaces
generated by a graph $(X,\arel)$, it is natural to consider distance
operators that are obtained by the \emph{shortest path distance} of a
weighted graph obtained by assigning weights to the arcs in
$\arel$. In \cite{NBCLM15}, such models are used for spatial logics
featuring a metric variant of the \emph{surrounded} operator of
\cite{CLLM14}.
Note however that in some cases other notions of distance can be more
appropriate. For example, \emph{sampling} an Euclidean space is often
done using a \emph{regular grid}, where points of a graph are arranged
on multiples of a chosen \emph{unit interval}, and connected by edges using some
notion of connectivity (e.g. in 2-dimensional space, one typically
uses four or eight neighbours per point). Shortest-path distance and
Euclidean distance obviously divert in this case (see \autoref{fig:EUCL} and \autoref{fig:DST} in \autoref{sec:app}), no matter how fine
is the grid or how many finite neighbours are chosen. Medical images
introduce some more complexity due to the fact that multi-dimensional
voxels frequently are \emph{anisotropic}, that is, the sizes of a
voxel in each of the multiple dimensions of the image are different.

\paragraph{Global model checking through distance transforms.}

The model checking algorithm implemented in \topochecker is a
\emph{global} one. That is, all points of the considered space or
space-time are examined, and those satisfying the considered formula are marked
by the algorithm. Even though this classical approach has drawbacks --
mostly related to the restriction to \emph{finite} models, and the
fact that large parts of a model need to be stored in central memory
-- it is very helpful in the case of medical image analysis, as logical
operators may take advantage of global analysis of the whole
space. One example where this is particularly useful is the
computation of distance formulas. This can be done using so-called
\emph{distance transforms}. 
The concept of distance transform comes from the area of topology and geometry in computer vision \cite{KKB96}. The idea is extensively used in modern image processing. Given a multi-dimensional image equipped with Euclidean distance, global spatial model checking of formulas
$\dist{e(z)} \phi$ can be done in linear time with respect to the
number of points of the space, assuming that the computation time of
$e(z)$ is negligible, and that the computation of $\phi$ is linear in
turn, which is true for the spatial logic \SLCS of
\cite{CLLM14}. Consider a multi-dimensional image. The outcome $S$ of computing the
truth value of $\phi$ on each point of a model is a binary (multi-dimensional) image. The binary value stored in each point corresponds
to the truth value of $\phi$ on that point. From this data, it is
possible to define a transformed image, called the \emph{distance
  transform} of $S$, such that in every point $x$, a value
$d_x \in \reals$ is stored. The value of $d_x$ is meant to correspond
to the minimum distance between $x$ and a point
satisfying $\phi$. There exist both exact and approximate algorithms for computing distance transforms,
that are usually classified by their asymptotic complexity,
computational efficiency, and possibility of parallel execution. In particular, there are effective
linear-time algorithms
\cite{MQR03,SURVEYDT}. In global model checking, one first computes the
distance transform, and then in one pass, for each $x$, the quantitative value of
$d_x$ can be replaced with the boolean value of $e(d_x)$. %Distance
%transforms can also be used in a straightforward way to implement in
%linear time satisfaction of the distance-based surrounded operator of
%\cite{NBCLM15}.
It is worth noting that similar algorithms exist for weighted graphs equipped with shortest-path distance. In this case, asymptotic complexity is generally speaking quasi-linear but not linear, although execution time is highly dependent on the structure of the considered graph.

In the next section we shall discuss how distance operators can be combined with texture analysis to identify tissues of different nature that lay within a certain distance of each other.

\section{Texture analysis operators}
\label{sec:ta}

Texture analysis (TA) operators are designed for finding
and analysing patterns in medical images, including some that are
imperceptible to the human visual system. Patterns in images are
entities characterised by brightness, colour, shape, size, etc.  TA
includes several techniques and has proved promising in a large number
of applications in the field of medical imaging
\cite{Kassner2010,Lopes2011,Castellano2004,Davnall2012}; in particular
it has been used in CAD applications
\cite{Woods2007,Han2014,Heinonen2009} and for classification or
segmentation of tissues or organs
\cite{Chen1989,Sharma2008,RodriguezGutierrez2013}.
In TA, image textures are usually characterised by estimating some
descriptors in terms of quantitative features. Typically, such
features fall into three general categories: syntactic, statistical,
and spectral \cite{Kassner2010}. Our preliminary experiments have been
mostly focused on statistical approaches to texture
analysis. Statistical methods consist of extracting a set of
statistics descriptors from the distributions of local features at
each voxel. In particular, we studied \emph{first order} statistical
methods, that are statistics based on the probability density function
of the intensity values of the voxels of parts, or the whole, of an
image, approximated as a histogram collecting such values into \emph{batches}
(driven by ranges). In this approach, the specific pixel adjacency
relationship is not taken into account. Common features are
statistical indicators such as \emph{mean}, \emph{variance},
\emph{skewness}, \emph{kurtosis}, \emph{entropy}
\cite{Srinivasan2008}.  Although a limitation of these operators is
that they ignore the relative spatial placement of voxels, statistical
operators are important for MI as their application is invariant under
transformations of the image. In particular, first order operators
are, by construction, invariant under \emph{affine transformations}
(rotation and scaling), which is necessary when analysing several
images acquired in different conditions. Nevertheless it is possible
to construct features using first order operators, keeping some
spatial coherence but losing at least partially the aforementioned
invariance \cite{Tijms2011}.

In our experimental evaluation, we defined a logical operator, called
SCMP -- for \emph{statistical comparison} -- that compares areas of an
image that are statistically similar to a predetermined area $SA$,
identified using a sub-formula. More precisely, SCMP is used to search
for sub-areas in the image whose empirical distribution is similar to
that of $SA$, up-to a user-specified threshold. For each voxel, a small surrounding area is considered; its statistical distribution is compared to that of $SA$ and a threshold is applied, obtaining a Boolean value that denotes whether the voxel belongs to an area statistically similar to $SA$. Statistical
distributions are compared using the \emph{cross-correlation}
function. Currently, the syntax and interpretation of the operator are
quite experimental, and different methods may be used to define the
data aggregation and comparison of distributions which are needed for
its implementation. Complexity is, among others, an issue to deal
with. %These aspects will be detailed in future work.
% However, we can
%already discuss the general functionality and features of such an
%operator when used for medical image analysis purposes.

\begin{figure}[!h]
  \centering
  \begin{subfigure}{0.4\textwidth}
    \includegraphics[width=\textwidth]{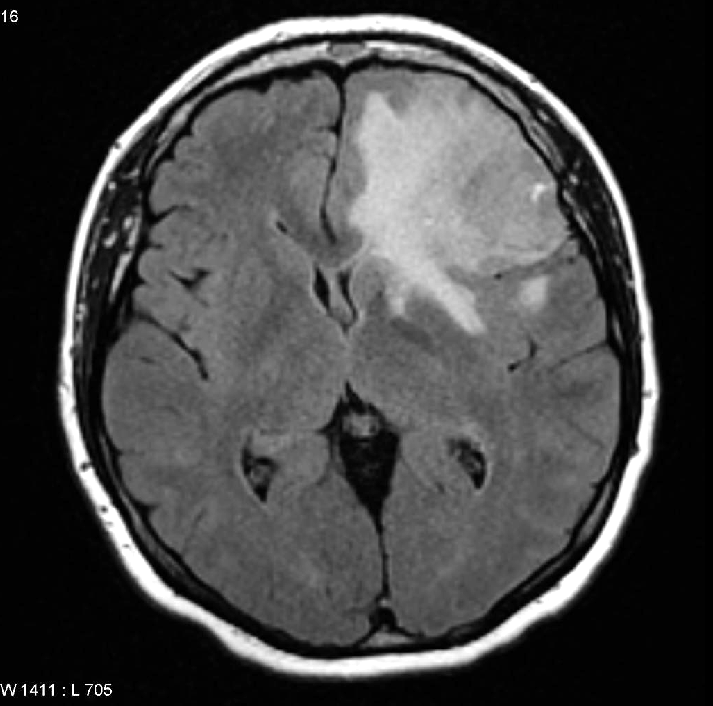}    
    \caption{A slice of a FLAIR MR acquisition of a brain affected by
      a glioblastoma. \phantom{Same slice superimposed to
        \emph{regions of interest} (ROIs) determined by threshold and
        distance operator. The yellow ROI is the oedema and the orange
        is the tumor.} ~ ~ ~ }
    \label{fig:GBM-Flair}
  \end{subfigure}~ ~ ~ 
  \begin{subfigure}{0.4\textwidth}
    \centering
    \includegraphics[width=\textwidth]{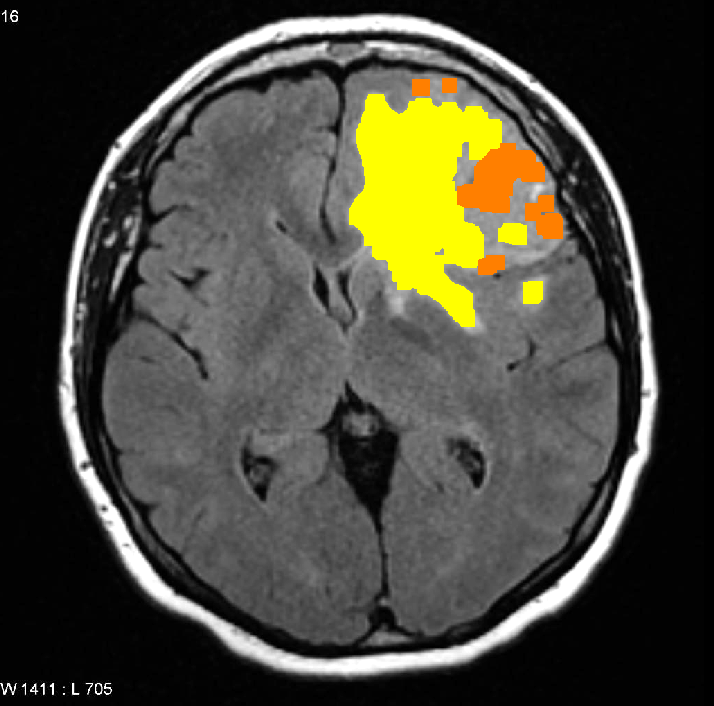}
    \caption{Same slice superimposed to \emph{regions of interest}
      (ROIs) determined by threshold and distance operator. The yellow
      ROI is the oedema; the orange one is the tumor.}
    \label{fig:GBM-soglia}
  \end{subfigure}
  \vspace{3mm}

  \begin{subfigure}{0.4\textwidth}
    \centering
    \includegraphics[width=\textwidth]{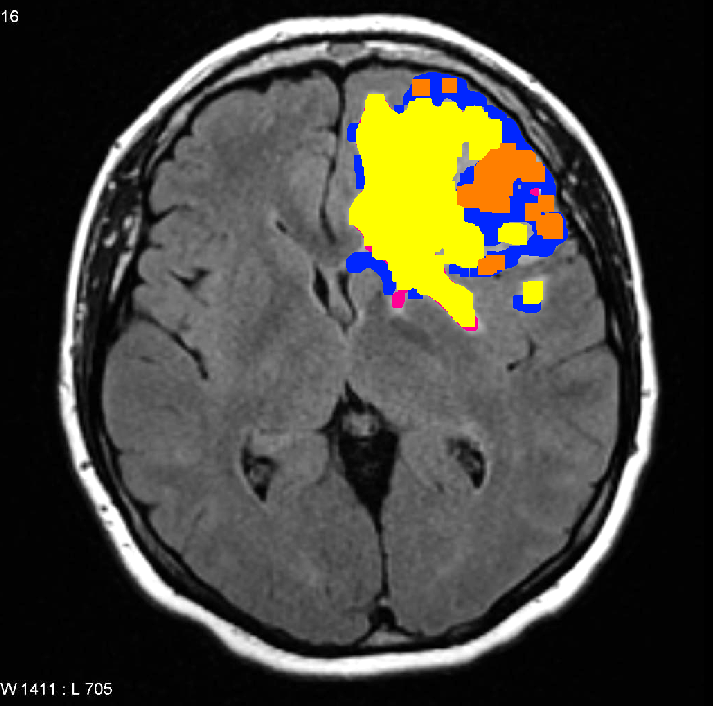}
    \caption{Same slice superimposed to ROIs determined by the SCMP
      operator, starting from orange and yellow ROIs. In blue, the additional tumor voxels; in magenta, the additional oedema
      voxels.}
    \label{fig:GBM-SCMP}
  \end{subfigure}~ ~ ~ 
  \begin{subfigure}{0.4\textwidth}
    \centering
    \includegraphics[width=\textwidth]{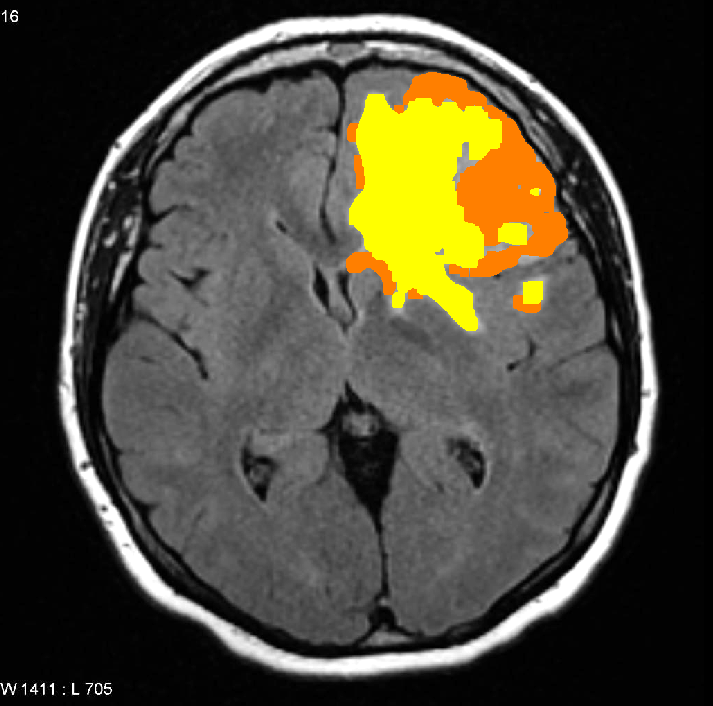}
    \caption{Final result of application of threshold, distance, and the SCMP operator. The yellow area is the identified oedema; the orange area
      is the tumor. \phantom{Same slice superimposed to ROIs
        determined by the SCMP operator. Blue are the tumor voxels and
        magenta the oedema voxels identified by SCMP starting from
        orange and yellow ROIs respectively.}}
    \label{fig:GBM-totale}
  \end{subfigure}
  \caption{Experimental results obtained by applying \topochecker to a medical image of a
    brain (case courtesy of A.Prof Frank Gaillard,
      Radiopaedia.org, rID: 5292).}
  \label{fig:GBM}
\end{figure}

The SCMP operator results in a generalisation of classical TA based on first-order statistics, since it analyses the statistical
distribution of a neighbourhood of each voxel \emph{as a whole}, whereas classical techniques for TA resort to the extraction of specific indicators. In \autoref{fig:GBM} we show the
output of \topochecker, enhanced with the SCMP operator, and
the distance-based operators described in \autoref{sec:dist}, applied to a slice of a MR-Flair acquisition
of a brain affected by glioblastoma (GBM), an intracranial
neoplasm\footnote{Case courtesy of A.Prof Frank Gaillard,
  Radiopaedia.org, rID: 5292}. GBMs are tumors composed of typically
poorly-marginated, diffusely infiltrating necrotic masses. Even if
the tumor is totally resected, it usually recurs, either near the
original site, or at more distant locations within the brain. GBMs are
localised to the cerebral hemispheres and grow quickly to
various sizes, from only a few centimetres, to lesions that cover a
whole hemisphere. Infiltration beyond the visible tumor margin is
always present. In MR T2/Flair images GBMs appear \emph{hyperintense}
and surrounded by \emph{vasogenic oedema}\footnote{Vasogenic oedema is an abnormal accumulation of fluid
from blood vessels, which is able to disrupt the blood brain barrier and invade
extracellular space}.

Being able to segment tumor and oedema in medical images can be of
immediate use for \emph{automatic contouring} applications in
radiotherapy and, in perspective, it can be helpful in detecting the
invisible infiltrations in CAD applications. In
\autoref{fig:GBM-soglia} we show a segmentation obtained using \topochecker. First, two
thresholds are applied to the original image, in order to identify two areas of the image
with particular brightness, that are supposed to loosely correspond to
an oedema and a tumor. Furthermore, the distance operator
(shortest path distance with 9-neighbours, see \autoref{sec:app}) is used to imposes a certain
degree of proximity between the oedema and the tumor. The voxels assigned to the oedema are drawn in yellow; the voxels
assigned to tumor are drawn in orange. These areas are used as two different $SA$ for the analysis shown
in \autoref{fig:GBM-SCMP}, obtained using the SCMP operator to further
enlarge the two regions, by searching areas that, although not falling
in the specified thresholds, are still identifiable as oedema and
tumor.  The model checker colours the additional voxels assigned to tumor in
blue; these areas are characterised by a statistical distribution
similar to the orange region. The tool colours the additional voxels
assigned to oedema in magenta. These are obtained by searching for areas having statistical distribution
similar to the yellow part of the image. In \autoref{fig:GBM-totale}
we show the final result of the procedure with tumor voxels in orange
and oedema voxels in yellow.

%!TEX root=main.tex
\section{Discussion}
We have just started the exploration of logical methods for medical
image analysis in the domain of radiotherapy. Logical properties are used as classifiers for points
of an image; this can be used both for colouring regions that may be
similar to diseased tissues, and therefore being diseased tissue in turn,
and for colouring regions corresponding to organs of the human
body. Envisaged applications range from \emph{contouring} to
\emph{computer-aided diagnosis}.
The field of Spatial Logics can benefit from such kinds of research;
for example, texture analysis operators may be defined, and in
particular operators that compare regions based on statistical
similarity. It will be interesting to study relevant axioms and
theories that can deal with the uncertainty generated by analysis of
statistical similarity in theorem proving or completeness studies, as
well as theories of bisimilarity or minimisation of models.
From the model-theoretical point of view, questions of interest relate
to the various kinds of distances that may arise in the considered
spatial models, ranging from classical Euclidean distance to
shortest-path distance on weighted graphs, on their axioms, and the
relation between different notions. The effect of approximate distance
transforms on the representational complexity of models may be an
interesting question for future work.
Our early experiments show that typical analyses carried out using spatial model checking in medical imaging require careful calibration of numeric parameters (for example, a threshold for the distance between a tumor and the associated oedema, or the size of areas identified by a formula, that are small enough to be considered \emph{noise}, and ought be filtered out). The calibration of such parameters can be done using machine-learning techniques. In this respect, future work could be focused on application, in the context of our research line, of the methodology developed by Bartocci et al. (see e.g. \cite{GBB14,BBMNS15}).
Some recent research focused on themes that are close to our planned development. In particular, \cite{SGBM16} uses spatio-temporal model checking techniques inspired by \cite{Gr+09} -- pursuing machine learning of the logical structure of image features -- to the detection of tumors. In contrast, our approach is more focused on human-intelligible logical descriptions. The work \cite{PG16} is closer to the setting of CAS, applied to biological processes, with an interesting focus on \emph{multi-scale} aspects.
The research line that we present in this paper stems from research in
collective adaptive systems and departs from it to direct spatial
analysis to medical imaging. We foresee that the novel statistical
texture analysis operators and the study of global model checking of
distance formulas using distance transforms are of interest when
dealing with very large populations that are spread over some spatial
structure (e.g. a geographical map). Potential applications include
the analysis of statistical properties arising from \emph{gossip
  protocols} and \emph{disease spreading} models, in which statistical
distribution of features in space appears to be relevant.

\paragraph{Acknowledgements} The authors wish to thank Marco Di Benedetto for suggesting the application of distance transforms to improve the complexity of model checking of formulas with distances, and the Medical Physics department of Azienda Ospedaliera Universitaria Senese (director: Fabrizio Banci Buonamici) for institutional support and encouragement in this research program.

\bibliographystyle{eptcs}
%{\small
\bibliography{generic}
%}
\appendix

\section{Comparison of metrics}\label{sec:app}

\begin{figure}
	\centering
	\includegraphics[height=3.7cm]{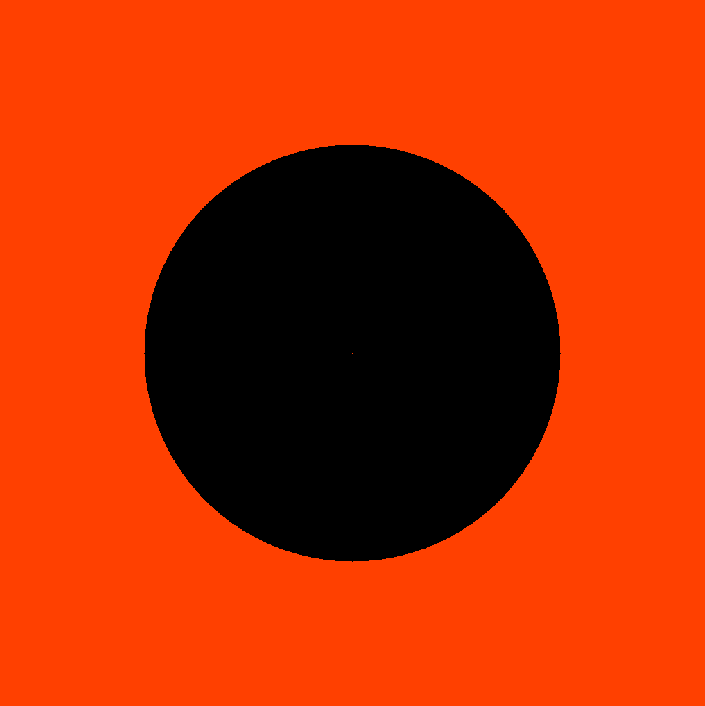}
	\caption{Error-free Euclidean distance (with threshold) from a
		point in the centre of image.}
	\label{fig:EUCL}
\end{figure}

\begin{figure}[!h]
	\centering
	\begin{subfigure}{0.4\textwidth}
		\centering
		\includegraphics[height=3.7cm]{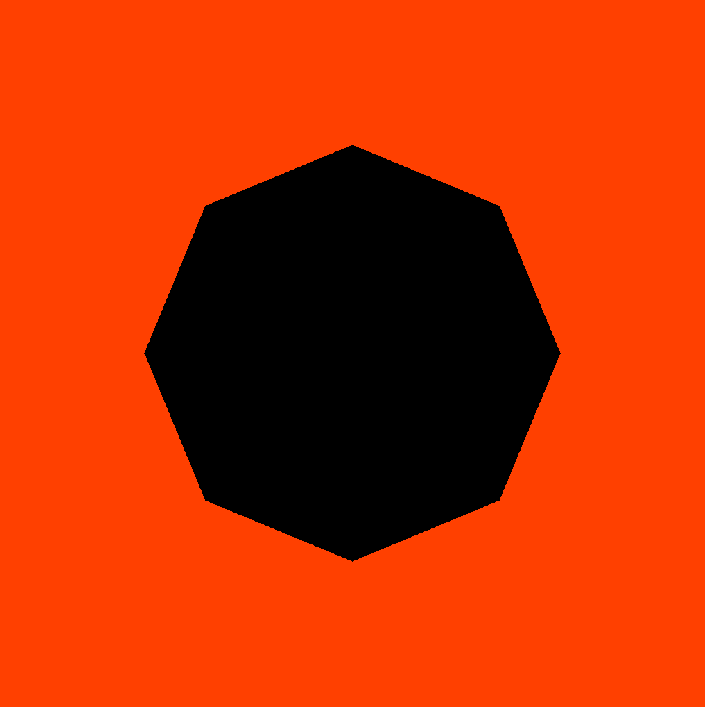}
		\caption{Chamfer distance in two dimensions with the Moore neighbourhood}
		\label{fig:MDDTw1}
	\end{subfigure}~ ~ ~ 
	\begin{subfigure}{0.4\textwidth}
		\centering
		\includegraphics[height=3.7cm]{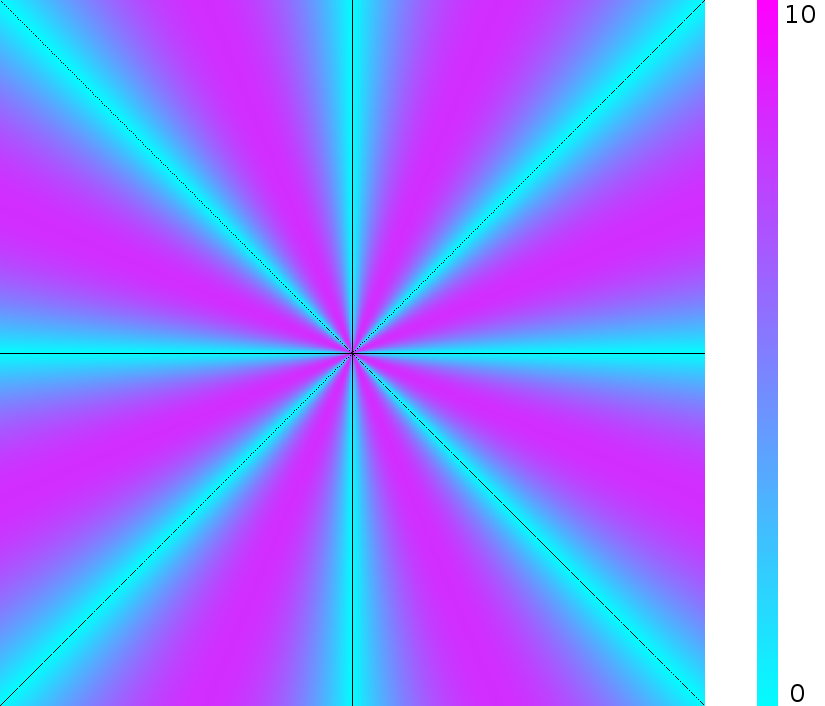}
		\caption{Percentage error with respect to the error-free Euclidean
			distance. Scale: 0-10\%}
		\label{fig:MDDTw1-err}
	\end{subfigure}
	\vspace{3mm}
	
	\begin{subfigure}{0.4\textwidth}
		\centering
		\includegraphics[height=3.7cm]{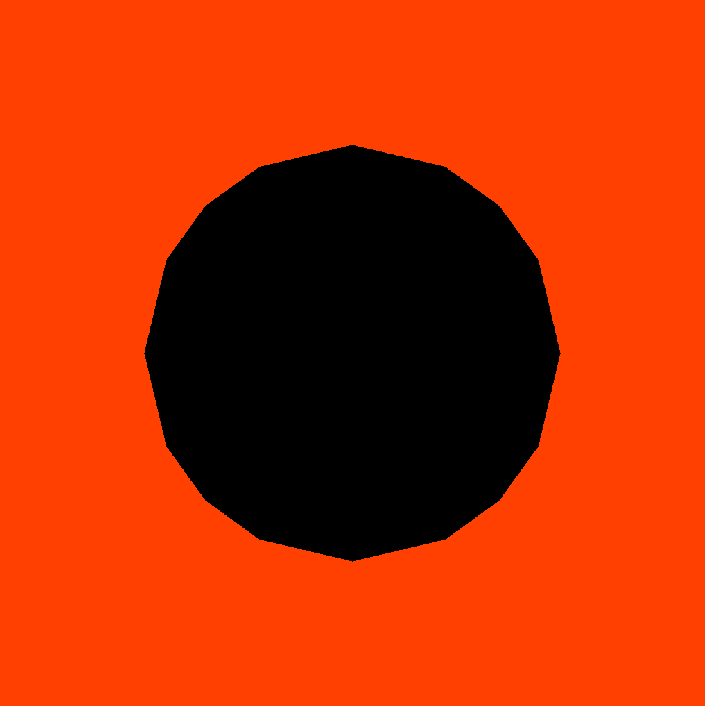}
		\caption{Chamfer distance in two dimensions with 24 chosen adjacent voxels.}
		\label{fig:MDDTw2}
	\end{subfigure}~ ~ ~ 
	\begin{subfigure}{0.4\textwidth}
		\centering
		\includegraphics[height=3.7cm]{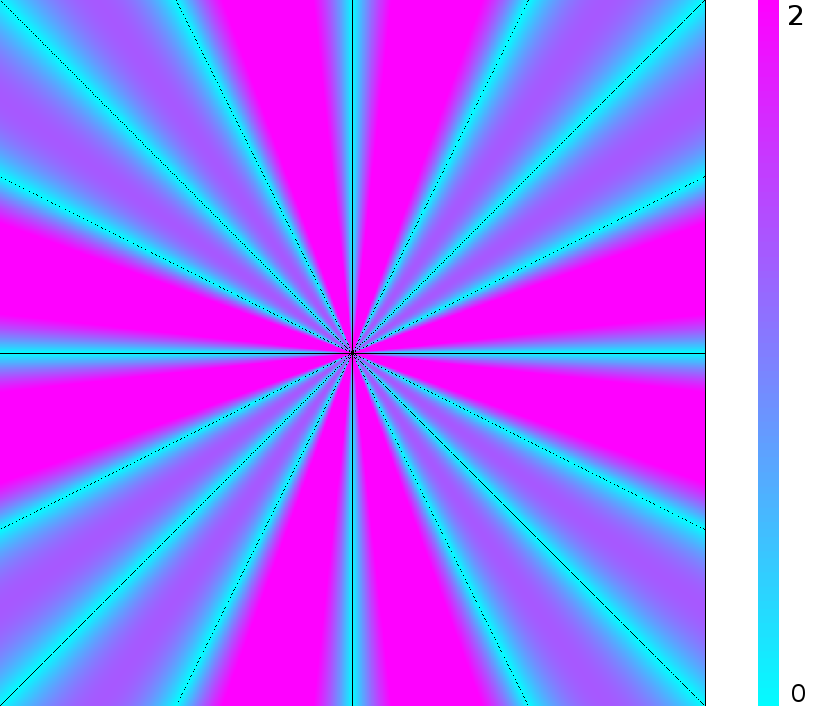}
		\caption{Percentage error with respect to the error-free Euclidean
			distance. Scale: 0-2\%}
		\label{fig:MDDTw2-err}
	\end{subfigure}
	% \vspace{3mm}
	
	% \begin{minipage}[l]{0.4\linewidth}
	%   ~ 
	% \end{minipage}
	% \begin{subfigure}{0.4\textwidth}
	%   \centering
	%   \includegraphics[width=0.8\textwidth]{cool-bar-h}
	% \end{subfigure}
	
	\caption{Distance operators, from the central point, with a threshold.}
	\label{fig:DST}
\end{figure}

In MI, Euclidean distance (based on the 2-norm) is the
reference distance between two voxels. Therefore, in this context, Euclidean
distance is considered \emph{error-free}. We shall now discuss different alternative definitions of distance that could be interesting for spatial model checking, both for medical imaging applications and also as a valuable addition to spatial analysis methods for CAS. 
Starting from \cite{Grevera2007}, we experimented with two distance transform operators,
the multi-dimensional Euclidean error-free distance operator, called EUCL in the following, and a distance
transform operator based on Dijkstra's shortest path algorithm, called MDDT (Modified Dijkstra Distance Transform),
that operates on the shortest-path distance of a graph constructed from the image. From the point of view of
the Dijkstra algorithm, an image is a graph whose vertices are the
voxels and whose arcs connect each voxel $x$ with chosen voxels that are considered adjacent to $x$. Every
arc is labelled with the chosen distance function applied to the two vertices that the arc
connects. In other words, we are considering a graph with nodes located in a distance space, arcs weighted according to the distance of the space, and a chosen notion of adjacency. In this particular case, the shortest-path distance is also called \emph{Chamfer distance}. The chosen adjacency is the most important factor in the precision-efficiency trade-off of the computed distance: the more adjacent voxels are considered, the more precise is the Chamfer distance when compared to the Euclidean distance, at the expenses of generating graphs with larger out-degrees. In \autoref{fig:EUCL} and \autoref{fig:DST} we show in red
the points satisfying $\dist{z> k} \phi$ for a binary image with only one point
satisfying $\phi$ (in the centre of the image).  In \autoref{fig:EUCL}
we show the output of the error-free EUCL operator. In
\autoref{fig:DST}, we show the output of the MDDT operator, alongside the characteristic pattern of the
percentage error with respect to Euclidean distance 
(\autoref{fig:MDDTw1-err} and \ref{fig:MDDTw2-err} -- see \cite{Grevera2007} for a detailed analysis of the percentage error of several distance transform algorithms). The percentage
error $\delta(x)$ for the distance transform $d(x)$ is defined in
every voxel $x$ as
$
\delta(x)=\frac{\left|d_{eucl}(x)-d(x)\right|}{d_{eucl}(x)}
$. 
%
% \todo{define all these metrics; next pargraph is unclear}
%
In \autoref{fig:MDDTw1} and \autoref{fig:MDDTw2} we show the
Chamfer distance obtained using MDDT, with different choices of adjacent voxels. In \autoref{fig:MDDTw1} the
adjacent voxels of a point are chosen to be its immediate neighbours on the main directions and diagonals (called \emph{Moore neighbourhood} in 2d images). In \autoref{fig:MDDTw2}, again the main directions and diagonals are used, but this process is iterated two times (that is, the chosen adjacent voxels of $x$ are all points in a hypercube of size 5, centred on $x$, except $x$ itself).
\enlargethispage{\baselineskip}

\begin{figure}
	\centering
	\begin{subfigure}{0.45\textwidth}
		\centering
		\includegraphics[height=3.7cm]{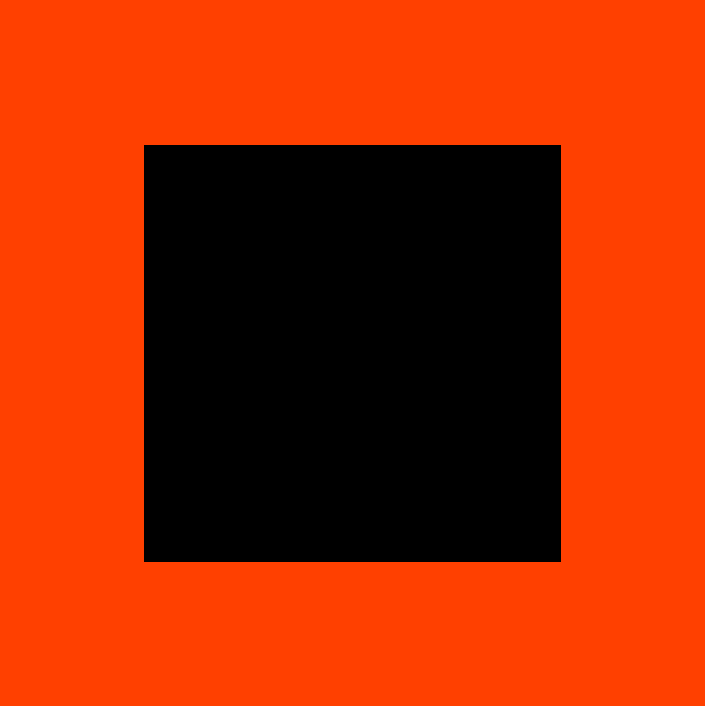}
		\caption{Chessboard distance: 
			$d_{\infty}(x,y)=\max_i{\left|x_i-y_i\right|}$}
		\label{fig:chessboard}
	\end{subfigure}
	\begin{subfigure}{0.45\textwidth}
		\centering
		\includegraphics[height=3.7cm]{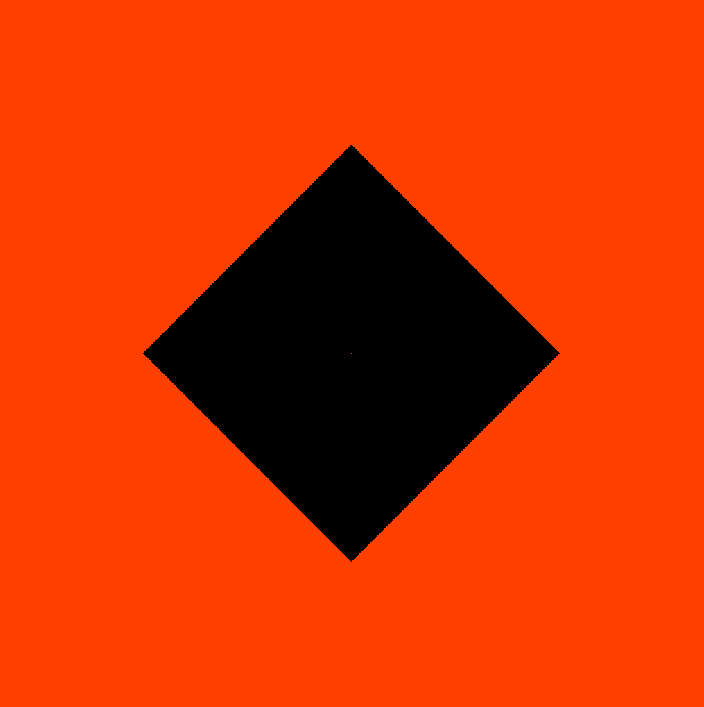}
		\caption{CityBlock distance:
			$d_{1}(x,y)=\sum_i{\left|x_i-y_i\right|}$ \phantom{aaa} } 
		\label{fig:cityblock}
	\end{subfigure}
	\caption{Non-Euclidean distance defined in an Euclidean space.}
	\label{fig:DSTNE}
\end{figure}

In a Euclidean space, Euclidean distance is not the only possible
distance; several definitions exist based on
different norms. In  (\autoref{fig:DSTNE}) we depict two widely used metrics.
In \autoref{fig:chessboard} adjacency is the same as in
\autoref{fig:MDDTw1} but all arcs have weight 1. This is called the
\emph{Chebyshev} or \emph{chessboard distance}, that in a Euclidean
space is the distance based on the so-called \emph{infinity-norm}. In
\autoref{fig:cityblock}, the underlying graph is as in \autoref{fig:chessboard}
but adjacency contains only voxels whose coordinates differ at
most in one dimension) (for 2D images, this is called \emph{Von Neumann neighbourhood}), thus obtaining the \emph{taxicab} or
\emph{cityblock distance}. In a Euclidean space, the cityblock
distance is the distance based on 1-norm.

\end{document}